\documentclass{elsart}
\usepackage{natbib}
\usepackage{graphicx}
\usepackage{amssymb}

\newcommand{\aap}{A\&A}
\newcommand{\apj}{ApJ}
\newcommand{\apjl}{ApJL}
\newcommand{\physrep}{Phys. Rev.}

\newcommand{\prd}{Phys. Rev. D}

\newcommand{\beq}{\begin{equation}}
\newcommand{\eeq}{\end{equation}}

\begin{document}

\begin{frontmatter}

\title{Abell 370: A Cluster with a Pronounced Triaxial Morphology}

\author[MIT]{E. De Filippis},
\ead{bdf@space.mit.edu}
\author[NA,NA2]{M. Sereno},
\ead{Mauro.Sereno@na.infn.it}
and
\author[MIT]{M.W. Bautz}
\ead{mwb@space.mit.edu}

\address[MIT]{Center for Space Research,
               Massachusetts Institute of Technology,
               70 Vassar Street, Building 37,
               Cambridge, MA 02139, USA}
\address[NA]{Dipartimento di Scienze Fisiche, Universit\`{a} degli Studi di Napoli ``Federico II'',
                Via Cinthia, Compl. Univ. Monte S. Angelo,
                80126 Naples, Italy}
\address[NA2]{INAF-OAC,
                Salita Moiariello, 16
                80131 Naples, Italy.}

\begin{abstract}
We have developed a new method to constrain the intrinsic three-dimensional shape of 
galaxy clusters using combined Sunyaev-Zel'dovich effect (SZE) and X-ray observations. 
We here present a first empirical implementation of this tool to an extremely 
intriguing distant galaxy cluster: Abell 370. 
\end{abstract}

\begin{keyword}
Galaxy clusters, A~370, X-rays \sep Sunyaev-Zel'dovich effect \sep cosmology, distance scale
\PACS 98.65.Cw \sep 95.85.Nv \sep 98.80.Es 
\end{keyword}

\end{frontmatter}

\section{Introduction}
\label{sec:Intro}
The intrinsic, three-dimensional (3-D) shape of clusters 
of galaxies is an
important cosmological probe. The structure of galaxy clusters is in
fact a helpful diagnostic of mass density in the universe and can
help in discriminating between different cosmological models. 
A realistic modelling of the intrinsic shape is hence critical for a
correct measure of the observed properties of clusters of galaxies.
Assumptions made on the cluster shape
strongly affect several other methods that estimate cosmological
parameters from observations of galaxy clusters: it in fact influences 
the determination of the Hubble constant and of the cluster mass and baryon
fraction~\citep{Coo98}. Also, projection effects can partially account
for the observed discrepancies in cluster mass determination between
lensing, X-ray and dynamical mass estimates.

\section{Theoretical Approach}
\label{sec:theoretical}
Cosmic microwave background (CMB) photons that pass through a cluster
interact with the energetic electrons of its hot intra-cluster medium (ICM) 
through inverse Compton scattering, with a probability $\tau\sim 0.01$. This
interaction causes a small distortion in the CMB spectrum, known as
the SZE~\citep{Sun70}. The SZE is
proportional to the electron pressure integrated along the line of
sight (l.o.s.), i.e. to the first power of the gas density. 
The measured
temperature decrement $\Delta T_{\rm SZ}$ of the CMB is:
\beq
\frac{\Delta T_{\rm SZ}}{T_{\rm CMB}} = f(\nu, T_{\rm e}) \frac{ \sigma_{\rm T} k_{\rm B} }{m_{\rm e} c^2}\int _{\rm l.o.s.}n_{\rm e} T_{\rm e} dl.
\label{eq:sze1}
\eeq
The cluster X-ray emission is due to bremsstrahlung from electron-ion
collisions; the X-ray surface brightness, $S_X$, is proportional to
the projection along the l.o.s of the square of the electron density:
\begin{equation}
\centering S_{\rm X} = \frac{1}{4 \pi (1+z)^4} \int _{\rm l.o.s.} n_{\rm e}^2 \Lambda_{\rm eH} dl.
\label{eq:sxb0}
\end{equation}
If we assume that the ICM is described by an isothermal triaxial
$\beta$-model distribution, then:
\beq
\label{eq:sz2}
\Delta T_{\rm SZ} = \Delta T_0 \left( 1+ \frac{\theta_{1}^2+e_{\rm proj}^2 \theta_{2}^2}{\theta_{\rm c,\rm proj}^2}\right)^{\frac{1}{2}-\frac{3\beta}{2}}; 
S_{\rm X} = S_{\rm X0} \left( 1+ \frac{\theta_{1}^2+e_{\rm proj}^2\theta_{2}^2}{\theta_{\rm c,\rm proj}^2} \right)^{\frac{1}{2}-3 \beta}
\label{eq:sxb1}
\end{equation}
where the central values are given by:
\beq
\label{eq:sze3}
\Delta T_0 \equiv T_{\rm CMB} f(\nu, T_{\rm e}) \frac{ \sigma_{\rm T} k_{\rm B} T_{\rm e}}{m_{\rm e} c^2} n_{\rm e0} \sqrt{\pi}\frac{\Gamma \left[3\beta/2-1/2 \right]}{\Gamma \left[3 \beta/2\right]}  \frac{D_{\rm c} \theta_{\rm c,proj}}{h^{3/4}}\sqrt{\frac{e_1 e_2}{e_{\rm proj}}} 
\eeq
\beq
\label{eq:sxb2}
S_{\rm X0} \equiv \frac{ \Lambda_{\rm eH} }{4 \sqrt{\pi} (1+z)^4} n_{\rm e0}^2 \frac{\Gamma \left[3\beta/2-1/2 \right]}{\Gamma \left[3 \beta/2\right]}  \frac{D_{\rm c} \theta_{\rm c,proj}}{h^{3/4}}\sqrt{\frac{e_1 e_2}{e_{\rm proj}}}
\eeq
In the above formulas: $T_{\rm e}$ is the temperature of the ICM, $k_{\rm B}$ the
Boltzmann constant, $T_{\rm CMB}$ the temperature of the
CMB, $\sigma_{\rm T}$ the Thompson cross section, $m_{\rm e}$ the
electron mass, $c$ the speed of light in vacuum, $D_{\rm c}$ the angular 
diameter distance to the cluster, $\theta_i$ 
the projected angular position on the plane of the sky (p.o.s.) of the observer 
orthogonal coordinate $x_{i,\rm obs}$, $h$ a function of the cluster shape and 
orientation~\citep{Sta77}, $e_{\rm i}$ the intrinsic axial ratios, $e_{\rm proj}$ the axial ratio 
of the major to the minor axes of the observed projected isophotes, 
$\theta_{c,\rm proj}$ the projection on the p.o.s. of the cluster angular 
intrinsic core radius, $\Lambda_{eH}$ the X-ray cooling function of the ICM in the
cluster rest frame and $f(\nu, T_{\rm e})$ 
accounts for frequency shift and relativistic corrections.\\
SZE and X-ray emission depend differently on the density of ICM, and
therefore also on the assumed cosmology. A joint analysis of SZE
measurements and X-ray imaging observations makes it then possible to
determine the distance to the cluster \citep{Bir99,Ree02}. One can
in fact solve Eqs.~(\ref{eq:sze3}) and (\ref{eq:sxb2}) for the
angular diameter distance $\left. D_{\rm c}\right|_{\rm Exp}$, by eliminating $n_{\rm e0}$.
If the redshift of a cluster is known, the angular diameter distance 
$\left. D_{\rm c}\right|_{\rm Cosm}$ between an observer and a source can be easily computed, for a fixed cosmology.\\
Under the assumption of spherical symmetry for the cluster, the experimental and the 
cosmological angular diameter distances assume the same value: 
\beq
\left. D_{\rm c}\right|_{\rm Cosm}=\left. D_{\rm c}\right|_{\rm Exp}^{\rm Sph}.
\label{eq:spherical}
\eeq 
Since under this assumption the 3-D morphology of the cluster is completely known, 
$\left. D_{\rm c}\right|_{\rm Exp}^{\rm Sph}$ can be computed combining
SZE and X-ray observations (Eqs.~\ref{eq:sze3}, \ref{eq:sxb2}).
The standard approach in the past decades has been to take advantage of this
complete knowledge of $\left. D_{\rm c}\right|_{\rm Exp}^{\rm Sph}$
under the assumption of spherical symmetry, 
to estimate $H_0$, provided that $\Omega_{M0}$ and $\Omega_{\Lambda 0}$ are
known from independent observations~\citep{Ree02,Mas01}.\\
The same approach clearly cannot be applied when the assumption of spherical symmetry is
relaxed and clusters are considered as more general triaxial systems. 
While $\left. D_{\rm c}\right|_{\rm Exp}$ is computed integrating cluster physical 
properties along the l.o.s., $\left. D_{\rm c}\right|_{\rm Cosm}$ is based on the cluster 
size on the p.o.s. If the cluster is not spherically symmetric, Eq.~\ref{eq:spherical} is 
therefore replaced by the more general:
\beq
\left. D_{\rm c}\right|_{\rm Cosm} = \left. D_{\rm c}\right|_{\rm Exp}^{\rm Sph}  h^{3/4} \left(\frac{e_{\rm proj} }{e_1 e_2} \right)^{1/2}.
\label{eq:dis2}
\eeq
The classical approach used in the past years can though be easily inverted 
to constrain the three-dimensional morphology of a cluster.
Here we follow this opposite approach.
We assume both the values of $\Omega_i$ and of $H_0$ to be known: 
$\left. D_{\rm c}\right|_{\rm Cosm}$ can be then determined. 
We then use Eq.~(\ref{eq:dis2}), together with other geometrical relations~\citep{DeF05}, 
to infer the three-dimensional morphology 
of a sample of galaxy clusters. We believe our assumption of a known cosmology to be fully justified by
recent results from observational cosmology, which have allowed an
unprecedented accuracy in testing theoretical models. 
Consistently with these results~\citep{Teg04} we use a flat
($\Omega_{\rm K}= 0$) cosmological model with
$H_0=71^{+4}_{-3}\ {\rm km\ s^{-1}\ Mpc^{-1}}$, $\Omega_{\rm M}=0.29 {\pm} 0.07$ and with a
non null cosmological constant.

\section{Empirical Test on A~370}
\label{sec:empirical}
\begin{figure}
\begin{center}
\includegraphics*[width=3.5cm]{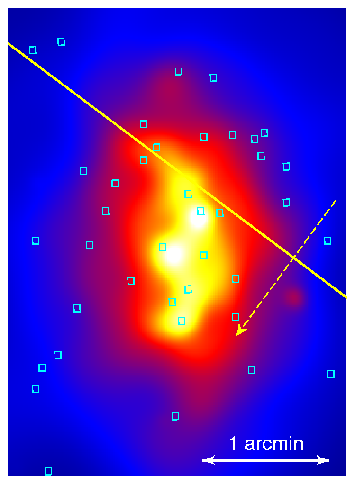}
\hspace{0.6cm}
\includegraphics*[width=3.5cm]{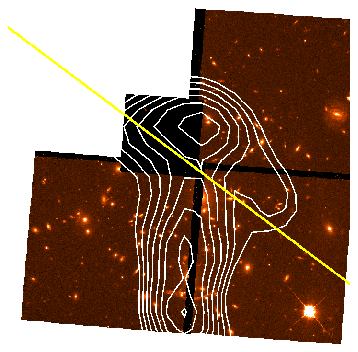}
\end{center}
\caption{{\it Left panel}: adaptively smoothed {\it Chandra} image of A~370 in the 0.3-7.0 keV energy band. Superposed squares give the positions of galaxies in the cluster redshift range. {\it Right panel}: {\it HST} image (scaled to the same physical size of the left panel). Superposed are the contours from the HR X-ray map; peaks point to increasingly softer ratios.}
\label{fig:observations}
\end{figure}

Abell 370 (z=0.375) is a rich cluster for which both strong and weak lensing have been detected. 
In the optical wave-band its emission is dominated by two giant elliptical galaxies. 
Both the optical and X-ray morphology are strongly elongated in the north-south direction. 
Lensing analyses have provided evidence that the cluster has a bimodal structure and that it 
is not fully relaxed~\citep{Bez99,Kne93}. 
The {\it Chandra} observation (left panel in Fig.~\ref{fig:observations}) shows a remarkably 
elliptical, diffuse X-ray emission.
Analyzing the redshift distribution of all cluster galaxies with spectroscopic redshifts, 
we have measured a trend of increasing redshift toward south-east (in the direction shown 
by the dashed arrow). 
In particular:
\vspace{-0.5cm}
\begin{itemize}
\item  galaxies north-west of the solid line are distributed in a stripe-like morphology 
parallel to the line -i.e. at $\sim 40^{\circ}$ respect to the main cluster elongation- and have an 
average redshift z=0.368;
\item  galaxies south-east of the line do not instead show a preferential direction and 
have an higher average redshift (z=0.382).
\end{itemize}
\vspace{-0.5cm}
Contours from an X-ray hardness ratio (HR) map of the cluster are superposed on its {\it HST} image, in the right panel in Fig.~\ref{fig:observations}. 
The HR map shows two soft peaks located in the south portion of the cluster, in coincidence
with two bright cluster galaxies. A third soft peak is located just north of the solid line: 
this is much more diffuse than the two southern ones and it is oriented in the same direction 
of the galaxy distribution.\\
Strong discrepancies have been observed for this cluster among the values of the total mass computed at different wavelengths; also average cluster temperatures estimated with different instruments strongly disagree. This comes as no surprise since such an irregular cluster, which contains substantial substructures, might not be fully virialized and partially supported by turbulent bulk gas motion; possible deviations from isothermality can also partially affect the cluster total mass determination. Although the hypotheses of hydrostatic equilibrium and isothermal gas are very strong, it has been recently shown that mass densities obtained under such assumptions
can yield accurate estimates even in dynamically active clusters, once the gas density distribution is properly mapped~\citep{DeF04,Ett03}.

\subsection{Results}
\label{sec:results}
\begin{figure}
\begin{center}
\includegraphics*[width=3.5cm]{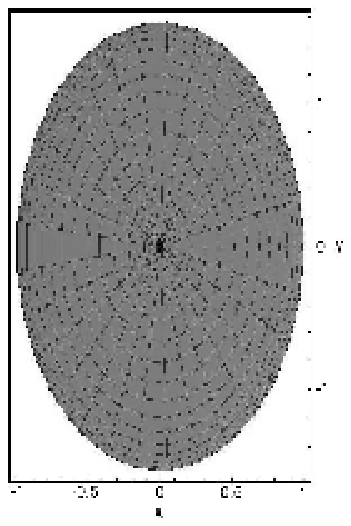}
\hspace{0.6cm}
\includegraphics*[width=3.5cm]{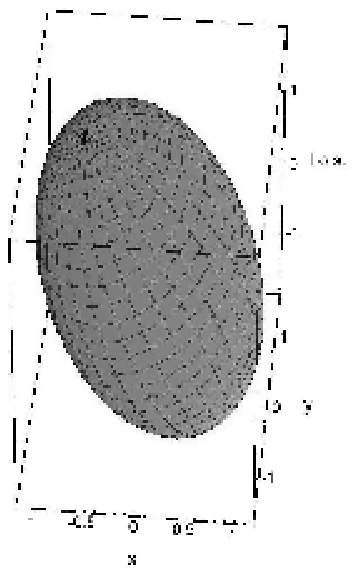}
\end{center}
\caption{{\it Left panel}: A~370 as it appears projected on the plane of the sky. {\it Right panel}: 3-D view of A~370 showing the cluster elongation along the l.o.s..}
\label{fig:math}
\end{figure}

A 370 is part of a sample of 25 clusters on which we are performing this combined X-ray and SZE analysis~\citep{DeF05}. The cluster shows an highly elliptical X-ray diffuse emission, with projected axes ratio: $e_{\rm proj}=1.56$. With our combined X/SZE analysis we estimate a value of the angular diameter distance: $\left. D_{\rm c}\right|_{\rm Exp}^{\rm Sph}=1221\ {\rm Mpc}$, while its cosmological value is: $\left. D_{\rm c}\right|_{\rm Cosm}=1063\ {\rm Mpc}$.\\
A comparison of these two estimates of the angular diameter distance reveals that the cluster is also strongly elongated along the l.o.s. (see Fig.~\ref{fig:math}) with a sharp triaxial morphology: $e_{\rm l.o.s.}=1.87$, where $e_{\rm l.o.s.}$ is the ratio between the cluster elongation along the l.o.s. and the minor axis in the p.o.s..\\
Our finding of a strong elongation along the l.o.s. and of a redshift difference between the two main galaxy concentrations, together with results from lensing which have provided evidence that the cluster has a bimodal structure, strongly suggest that A~370 has two (or more) substructures in the process of merging, which must be taking place at a small angle respect to the l.o.s. A~370 turns out to be one of the clusters in our sample with the most strongly pronounced triaxial morphology.

\section{Summary and Future Work}
\label{sec:summary}
We have applied a combined X-ray/SZE analysis to the galaxy cluster A~370 in order to infer its triaxial morphology. A~370 turns out to be strongly elongated along the l.o.s., with two (or more) substructures in the process of merging. Spectroscopic redshifts of member galaxies suggest that the process must be taking place at a small angle respect to the l.o.s..\\
With such a large merger taking place we expect gas bulk motions of $\sim 2000\ {\rm km/sec}$ in the cluster; this will lead to a shift in the center of the emission line $>40\ {\rm  eV}$. Possible broadening of emission lines of heavy elements, due to hydrodynamic turbulence in the IC gas would also be present. Since both the above features can be easily detected by {\it XRS}, A370 would be a perfect target for {\it ASTRO-E2}. This would lead to a possible measurement of bulk motion of the IC gas and allow to constrain the dynamical stages of the ICM.

\vspace{0.1cm}
This work was supported by NASA grants NAS8-39073 and NAS8-00128.

\end{document}